\documentclass[reprint,amsmath,amssymb,aps,]{revtex4-2}

\usepackage{graphicx}
\usepackage{dcolumn}
\usepackage{bm}
\usepackage[]{hyperref}
\usepackage{color} 
\usepackage{amsmath}
\usepackage{mathrsfs}
\usepackage{hyperref} 
\usepackage{ulem} 
\hypersetup{
	colorlinks=true,
	linkcolor=blue,
	filecolor=blue,      
	citecolor=blue,
	urlcolor=blue  
}

\begin{document}

\title{Polarization Entanglement in Atomic Biphotons via OAM-to-Spin Mapping}

\author{Chang-Wei Lin,$^1$ Yi-Ting Ma,$^1$ Jiun-Shiuan Shiu,$^{1,2}$ and Yong-Fan Chen$^{1,2}$}

\email{yfchen@mail.ncku.edu.tw}

\affiliation{
$^1$Department of Physics, National Cheng Kung University, Tainan 70101, Taiwan\\ 
$^2$Center for Quantum Frontiers of Research $\&$ Technology, Tainan 70101, Taiwan
}

\date{December 12, 2025}

\begin{abstract}

We demonstrate polarization-entangled biphotons in a cold-atom double-$\Lambda$ system, overcoming atomic selection rules that suppress polarization correlations and favor orbital angular momentum (OAM) entanglement. Using spatial light modulators, we coherently map a selected two-dimensional OAM subspace onto the polarization basis and thereby open an otherwise inaccessible polarization channel. Quantum-state tomography confirms that the mapping preserves the biphoton coherence. The four polarization Bell states are generated with fidelities of $92\text{--}94\%$ with few-percent statistical uncertainties, and an average Clauser--Horne--Shimony--Holt parameter of $S=2.44$ verifies the survival of nonlocal correlations. To the best of our knowledge, this work presents the first demonstration of OAM-to-polarization entanglement transfer in a cold-atom spontaneous four-wave mixing platform and establishes a practical interface for integrating atomic OAM resources with polarization-based quantum communication networks.

\end{abstract}


\maketitle


\newcommand{\FigOne}{
    \begin{figure*}[t]
    \centering
    \includegraphics[width = 17.2 cm]{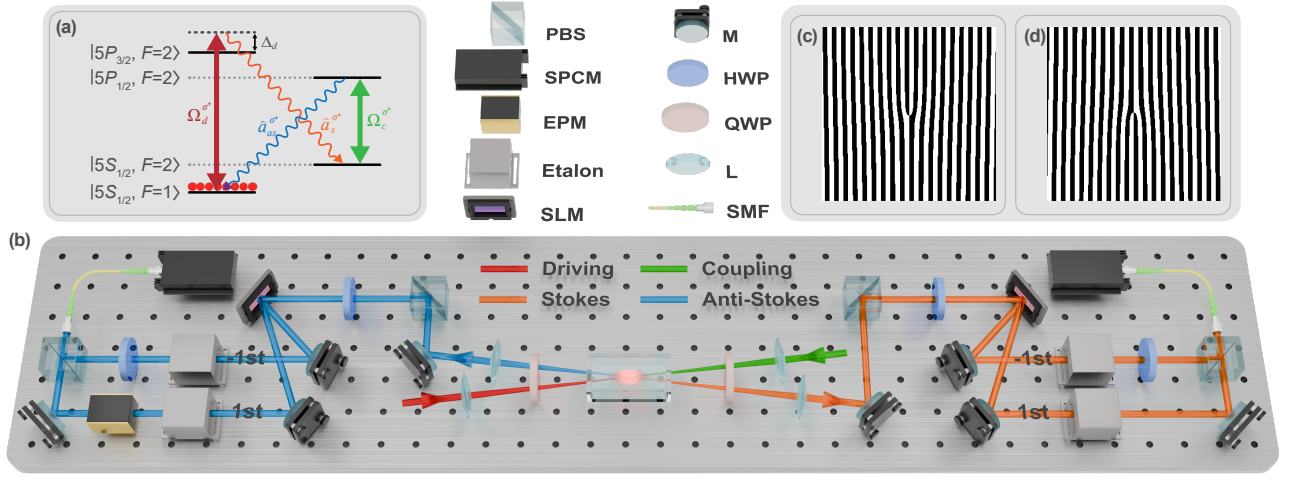}
    \caption{
    (a) Energy-level diagram of the $^{87}$Rb double-$\Lambda$ configuration used for SFWM biphoton generation.  
    (b) Experimental setup of the biphoton source and the OAM-to-polarization interface.  
    M: mirror; HWP: half-wave plate; QWP: quarter-wave plate; L: lens; SMF: single-mode fiber;  
    PBS: polarizing beam splitter; SPCM: single-photon counting module;  
    EPM: electro-optic phase modulator; SLM: spatial light modulator.  
    (c), (d) Computer-generated fork holograms displayed on the Stokes and anti-Stokes SLMs, which impart $\pm1$ units of OAM and define the two-dimensional OAM subspace used for mapping.
    }
    \label{fig:Setup}
    \end{figure*}
}
\newcommand{\FigTwo}{
    \begin{figure*}[t]
    \centering
    \includegraphics[width = 17.6 cm]{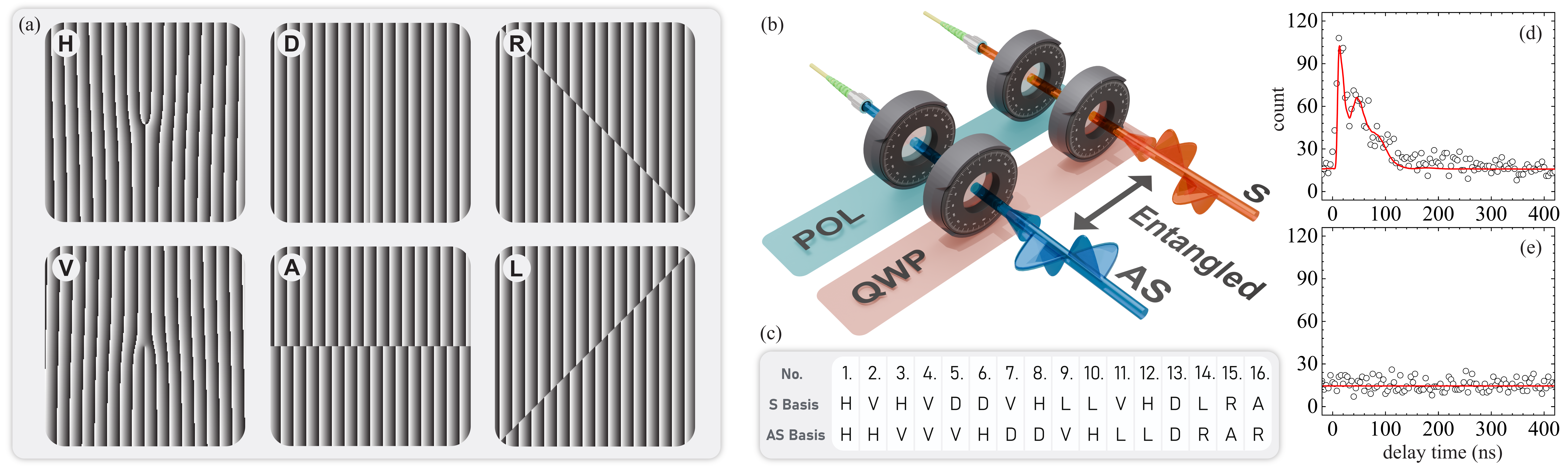}
    \caption{
    (a) Six computer-generated holograms used for OAM quantum-state tomography. The labels identify the measurement bases corresponding to the sixteen joint projection settings in panel (c). (b) Experimental configuration for tomographic projection. Each channel contains a quarter-wave plate (QWP) and a linear polarizer (POL), which project the biphoton state onto the six polarization bases defined by the orientations (POL, QWP): $|H\rangle$ $(0^\circ,0^\circ)$, $|V\rangle$ $(90^\circ,0^\circ)$, $|D\rangle$ $(45^\circ,0^\circ)$, $|A\rangle$ $(135^\circ,0^\circ)$, $|R\rangle$ $(0^\circ,45^\circ)$, and $|L\rangle$ $(90^\circ,45^\circ)$. (c) Sixteen joint measurement settings used for reconstructing the biphoton density matrix. (d), (e) Temporal cross-correlation functions of the biphotons. Panel (d) shows correlated projections (HH, VV) for the $|\Phi^{+}\rangle$ state, revealing the narrowband biphoton wave packet. Panel (e) shows the uncorrelated projections (VH, HV), where only accidental background remains. Black circles denote experimental data and red curves represent theoretical predictions.
    }
    \label{fig:Tomography}
    \end{figure*}
}
\newcommand{\FigThree}{
    \begin{figure}[b]
    \centering
    \includegraphics[width = 8.6 cm]{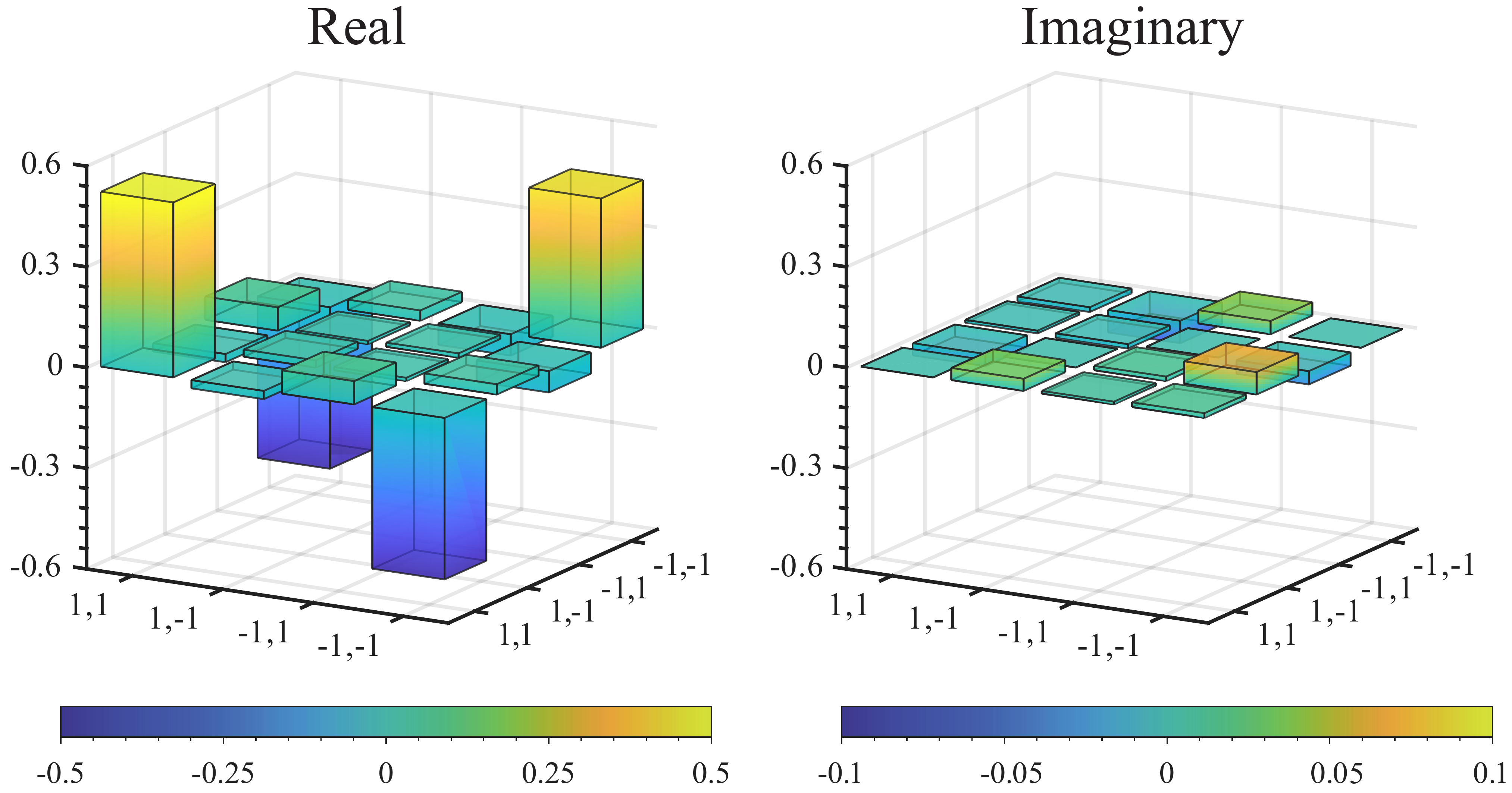}
    \caption{
    Reconstructed density matrix of the OAM-entangled biphoton state $|\Phi^{-}\rangle$ obtained with one SLM
    and one etalon in each channel.  
    Basis labels such as ``1,1'' denote the OAM state $|l_s=1,l_{as}=1\rangle_L$.  
    The left (right) column shows $\mathrm{Re}(\rho)$ [$\mathrm{Im}(\rho)$].
    }
    \label{fig:OAM}
    \end{figure}
}
\newcommand{\FigFour}{
    \begin{figure}[t]
    \centering
    \includegraphics[width = 8.6 cm]{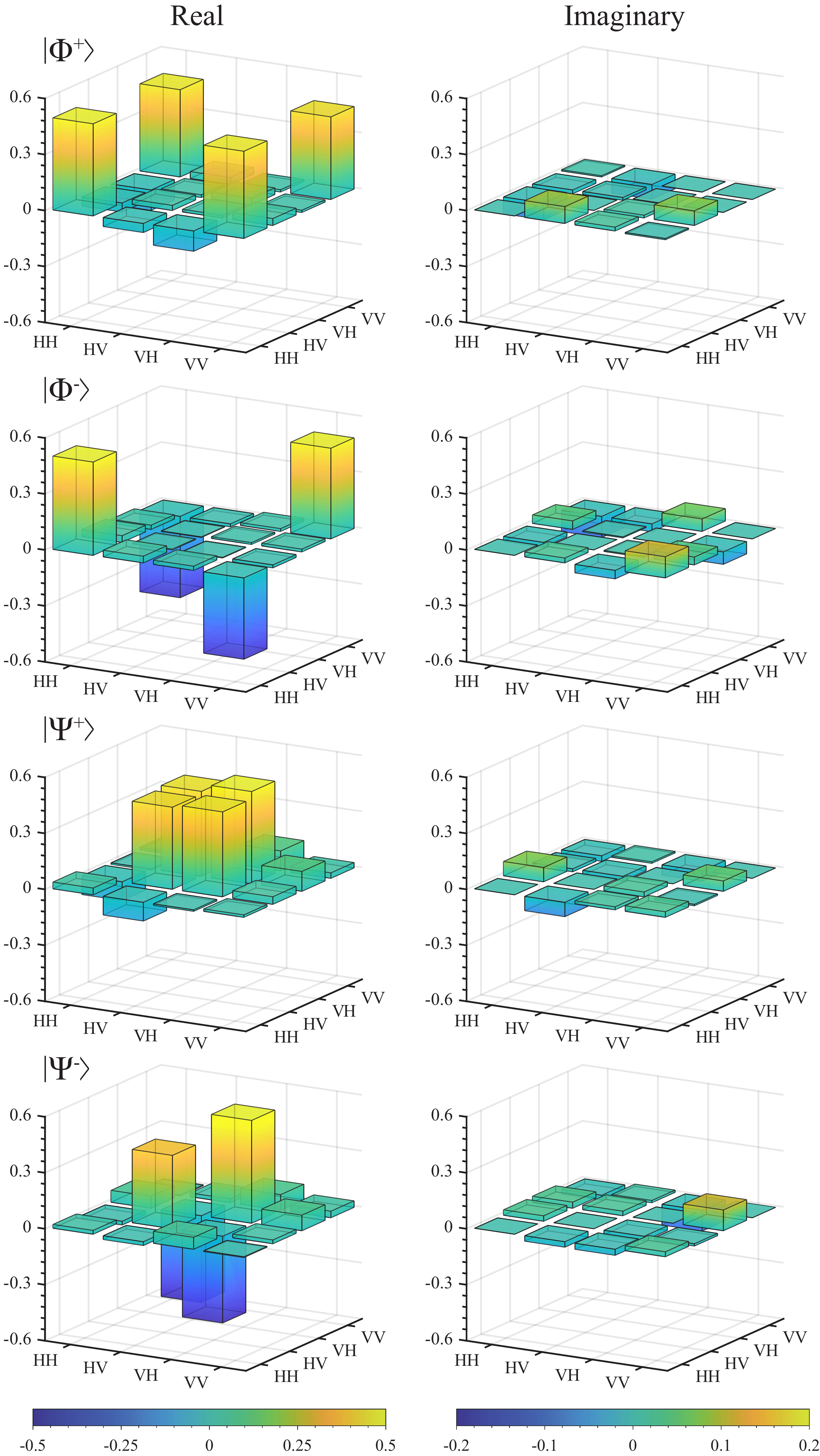}
    \caption{
    Reconstructed density matrices of the four polarization Bell states obtained using
    maximum-likelihood estimation.  
    Rows correspond to $|\Phi^{+}\rangle$, $|\Phi^{-}\rangle$, $|\Psi^{+}\rangle$, and $|\Psi^{-}\rangle$.  
    The left (right) column shows $\mathrm{Re}(\rho)$ [$\mathrm{Im}(\rho)$].  
    These reconstructions use two SLMs and four etalons, the configuration relevant for
    evaluating the OAM-to-polarization mapping fidelity.
    }
    \label{fig:Polarization}
    \end{figure}
}



Entangled photon pairs are a central resource for quantum communication protocols, including quantum key distribution (QKD) \cite{QKD1,QKD2}, quantum teleportation \cite{QT1}, and quantum repeaters for long-distance networks \cite{QR1,QR2,network1}. They can be generated in a variety of platforms, notably spontaneous parametric down-conversion (SPDC) in nonlinear crystals \cite{SPDC2} and spontaneous four-wave mixing (SFWM) in atomic media \cite{SFWM4,SFWM6,SFWM5,Ourtheory}. SFWM in cold atomic ensembles is particularly attractive because it provides narrowband, time-correlated biphotons that are intrinsically compatible with atomic transitions and quantum memories \cite{Memory1,Memory4}. Owing to angular momentum conservation, such biphotons naturally exhibit well-defined orbital angular momentum (OAM) correlations \cite{OAM1}, which make cold-atom SFWM a promising interface between atomic quantum nodes and photonic channels.

The OAM of light spans a discrete and, in principle, unbounded Hilbert space \cite{OAM2}, enabling high-dimensional quantum information processing and spatial-mode multiplexing \cite{OAM5,Protocol2}. As a complementary degree of freedom, the spin angular momentum of light, which manifests as polarization, forms a natural two-dimensional qubit basis that is widely used in quantum communication. However, distributing OAM-encoded entanglement over long distances is challenging. Standard single-mode fibers do not support OAM eigenmodes, and specialty fibers with tailored geometries improve performance at the cost of fabrication complexity and increased sensitivity to mode coupling and environmental perturbations \cite{OAMfiber2}. In practice, long-distance quantum links therefore overwhelmingly adopt polarization as the encoding basis.

In cold-atom SFWM systems, the generation of polarization-entangled biphotons is strongly constrained. In a standard double-$\Lambda$ configuration, atomic selection rules suppress polarization correlations, so the emitted photon pairs exhibit entanglement only in the OAM degree of freedom. Related OAM-to-polarization transfer has recently been demonstrated in a cavity-enhanced SPDC source \cite{SPDC4}, illustrating the usefulness of external mode-conversion interfaces. Here we demonstrate an analogous capability in a narrowband cold-atom platform that is naturally compatible with quantum memories and long-lived atomic quantum nodes. 

Our scheme relies entirely on external optical manipulation in contrast to methods that generate polarization entanglement by modifying the atomic level structure \cite{OtherMethod1} or pump geometry \cite{OtherMethod2}. Using spatial light modulators (SLMs) as coherent mode converters, we map the biphoton OAM entanglement onto the polarization degree of freedom within a selected two-dimensional OAM subspace. This approach circumvents the intrinsic selection-rule limitations of the double-$\Lambda$ system and opens an otherwise inaccessible polarization-entanglement channel. Combined with electro-optic phase control and digital holography, the interface enables flexible generation of polarization-entangled states while preserving the narrow bandwidth and memory-compatible characteristics of atomic biphotons.

Using this interface, we convert OAM entanglement into polarization entanglement with high fidelity and generate all four Bell states through active phase control. Quantum state tomography yields an average Bell-state fidelity of $93\%$, comparable to the highest-quality entangled-photon sources in atomic systems \cite{OtherMethod1,Otherteam1}. These results demonstrate a robust and noninvasive approach for integrating narrowband atomic OAM resources into polarization-based fiber networks and establish a versatile entanglement-transfer capability for hybrid quantum networking architectures.

\FigOne



The biphoton source is implemented in a cold $^{87}\text{Rb}$ ensemble prepared in a magneto-optical trap (MOT) and optically pumped into the $|5S_{1/2},F=1\rangle$ ground state [Fig.~\ref{fig:Setup}(a)]. SFWM is driven through a double-$\Lambda$ configuration using two counter-propagating laser fields. The $\sigma^{+}$ driving beam ($\Omega_d=2\Gamma$, $30~\mu\text{W}$) is detuned by $10\Gamma$ from the $|5S_{1/2},F=1\rangle \rightarrow |5P_{3/2},F=2\rangle$ transition and generates the Stokes field $\hat{a}_s$. The $\sigma^{+}$ coupling beam ($\Omega_c=4\Gamma$, $178~\mu\text{W}$) is tuned to the $|5S_{1/2},F=2\rangle \rightarrow |5P_{1/2},F=2\rangle$ transition and produces the anti-Stokes field $\hat{a}_{as}$. Operating at an optical depth of $48\pm2$, the source yields a biphoton generation rate of $5\times10^6~\text{s}^{-1}$ at the output of the atomic ensemble. An elongated atomic geometry enhances backward scattering into well-defined spatial modes. The Stokes and anti-Stokes photons are collected in opposite directions, and the collection axis is aligned at an angle of $1.7^{\circ}$ relative to both pump beams to reduce leakage from the strong driving and coupling fields. Each detection channel includes a polarization filter that selects the $\sigma^{+}$ component. The experimental cycle consists of a $5$-ms sequence containing ten driving pulses of $5~\mu\text{s}$ duration, separated by $15~\mu\text{s}$ intervals during which an auxiliary trapping pulse repumps atoms to the initial ground state. Additional information regarding the atomic configuration is available in Ref.~\cite{Detail}.

After emerging from the cold-atom medium, the Stokes and anti-Stokes photons pass through a quarter-wave plate (QWP) and a polarizing beam splitter (PBS) that jointly ensure selection of the $\sigma^{+}$ polarization component [Fig.~\ref{fig:Setup}(b)]. The photons are then directed onto an SLM (HOLOEYE PLUTO-2.1-NIR-113B) programmed with computer-generated holograms [Figs.~\ref{fig:Setup}(c) and \ref{fig:Setup}(d)] that function as forked diffraction gratings \cite{Hologram}. The first-order diffracted beams with OAM indices $+1$ and $-1$ are individually coupled into temperature-stabilized etalons to suppress residual background photons outside the biphoton bandwidth. The beam in the $-1$ order subsequently passes through a half-wave plate (HWP) that rotates its linear polarization by $90^{\circ}$. The two beams are then recombined on a PBS to form the polarization-entangled biphoton state. Further details concerning the computer-generated holograms are given in the Supplemental Material.


The OAM of light is encoded in the transverse spatial profile and is conveniently described in the Laguerre--Gaussian basis $\mathrm{LG}^l_p$ \cite{OAM2}, where $l$ denotes the azimuthal index and $p$ the radial index. Modes with distinct $l$ values are orthogonal and span a discrete Hilbert space for the transverse wavefunction.

\FigTwo

With both pump fields $\Omega_d$ and $\Omega_c$ carrying zero OAM, angular-momentum conservation constrains the SFWM-generated biphotons. In a co-propagating geometry this condition yields $l_s+l_{as}=0$, whereas in our counter-propagating configuration the opposite propagation directions require $l_s=l_{as}$. The biphoton state can thus be written as
\begin{equation}
	|\psi\rangle = c_0 |0,0\rangle_L + \sum_{l\ge1} c_l \bigl(|l,l\rangle_L + |-l,-l\rangle_L\bigr),
\end{equation}
with $c_l$ determined by spatial-overlap integrals and the Rayleigh lengths of the interacting modes \cite{Simpletheory}. In practice, the state is dominated by the $l=0$ and $\pm1$ components, which define the two-dimensional OAM subspace used for mapping.

To transfer this OAM entanglement to polarization, each photon is directed onto an SLM displaying a fork hologram [Fig.~\ref{fig:Setup}(c)], where the $n$th diffraction order imparts an OAM shift $l\rightarrow l+n$. A binary phase grating concentrates the efficiency into the $n=\pm1$ orders. After diffraction, each order is independently filtered by an etalon transmitting only the Gaussian mode $\mathrm{LG}^{0}_{0}$, thereby removing all nonzero-$l$ contributions.

Considering the dominant $l=0,\pm1$ components, the diffracted states for selected order pairs $(n_s,n_{as})$ are
\begin{align}
	(+1,+1) &: |1,1\rangle_L 
	+ |2,2\rangle_L + |0,0\rangle_L, \nonumber\\[3pt]
	(+1,-1) &: |1,-1\rangle_L 
	+ |2,0\rangle_L + |0,-2\rangle_L, \nonumber\\[3pt]
	(-1,+1) &: |-1,1\rangle_L 
	+ |0,2\rangle_L + |-2,0\rangle_L, \nonumber\\[3pt]
	(-1,-1) &: |-1,-1\rangle_L 
	+ |0,0\rangle_L + |-2,-2\rangle_L.
\end{align}
After etalon filtering, only the $l=0$ terms survive. Thus the $(+1,+1)$ and $(-1,-1)$ selections isolate the outputs corresponding to $|-1,-1\rangle_L$ and $|1,1\rangle_L$, respectively. These two outcomes define the effective two-dimensional OAM subspace $\{|1,1\rangle_L, |-1,-1\rangle_L\}$ used for the subsequent OAM-to-polarization mapping. In this sense, the combination of fork holograms and Gaussian-mode etalon filtering implements an effective projective selection of this subspace, which is subsequently mapped onto the polarization-qubit basis.



To synthesize the final polarization-entangled state, the $-1$ diffraction order is sent through an HWP that rotates its linear polarization so that it becomes orthogonal to the $+1$ order. The two paths are then coherently recombined at a PBS, which maps the OAM superposition $|1,1\rangle_L + |-1,-1\rangle_L$ onto the polarization basis and produces the state $|\psi\rangle = (|H,H\rangle_P + e^{i\theta}|V,V\rangle_P)/\sqrt{2}$. Here $\theta$ is the relative phase between the two components, controlled by an electro-optic phase modulator (EPM) placed in one of the diffraction paths. This control enables switching between $|\Phi^{+}\rangle$ and $|\Phi^{-}\rangle$. 

To access the complementary subspace corresponding to $|\Psi^{+}\rangle$ and $|\Psi^{-}\rangle$, we digitally rotate the hologram on one of the SLMs by $180^{\circ}$. This rotation reverses the sign of the imparted OAM and transforms the state to $|\psi\rangle = (|H,V\rangle_P + e^{i\theta}|V,H\rangle_P)/\sqrt{2}$. Because only the $l=\pm1$ components are converted into the Gaussian mode by the selected diffraction orders and subsequent spatial filtering, all other $l$ components are rejected by the etalons. The PBS therefore performs a coherent mapping on this two-dimensional OAM subspace and yields a well-defined polarization-entangled state.

To quantify the performance of the OAM-to-polarization interface, we perform full quantum-state tomography on both the native OAM-entangled state and the post-conversion polarization state \cite{Tomography,Tomography2}. Although the physical projection bases differ, the reconstruction protocol is identical and relies on sixteen projective measurements. For the OAM sector, the projections are implemented by programming six computer-generated holograms on the SLMs in the Stokes and anti-Stokes channels [Fig.~\ref{fig:Tomography}(a)]. For the polarization sector, the projections are realized using a QWP and a linear polarizer positioned before the collection fibers [Fig.~\ref{fig:Tomography}(b)]. The sixteen measurement settings in Fig.~\ref{fig:Tomography}(c) form a tomographically complete set, and the corresponding coincidence counts are used to reconstruct the density matrix $\rho_{\mathrm{exp}}$ by maximum-likelihood estimation \cite{Likelihood}. The fidelity $F$ with respect to a target Bell state $\rho_{\mathrm{tar}}$ is defined as
\begin{equation}
	F = \left[ \mathrm{Tr}\left( \sqrt{\sqrt{\rho_{\text{tar}}} \rho_{\text{exp}} \sqrt{\rho_{\text{tar}}}} \right) \right]^2,
	\label{eq(8)}
\end{equation}
which quantifies the overlap between the generated state and the ideal Bell state \cite{Fidelity}. Statistical uncertainties quoted below are obtained from a Monte Carlo resampling of the measured coincidence histograms assuming Poissonian counting statistics (see Supplemental Material). Here and throughout, numbers in parentheses denote one-standard-deviation statistical uncertainties. A detailed description of the tomographic reconstruction and the mapping between coincidence counts and projection probabilities is also provided in the Supplemental Material.

\FigThree



We first characterize the OAM entanglement of the biphoton source using one SLM and one etalon in each of the Stokes and anti-Stokes channels. The reconstructed density matrix in Fig.~\ref{fig:OAM} is dominated by the coherent superposition $|1,1\rangle_L - |-1,-1\rangle_L$. The measured fidelity with respect to $|\Phi^{-}\rangle_L$ is $96.4(2.2)\%$, obtained using Eq.~(\ref{eq(8)}). Because this value already incorporates the mode-projection limitations imposed by the two etalons, it should be regarded as a lower bound on the intrinsic OAM coherence produced by the atomic SFWM source.

\FigFour

We next evaluate the OAM-to-polarization interface by generating all four polarization Bell states and performing full tomography, as shown in Fig.~\ref{fig:Polarization}. In this configuration, each SLM generates spatially separated $\pm1$ diffraction orders, and each order is filtered by an independent etalon. The polarization tomography therefore employs two SLMs and four etalons, which doubles the number of filtering stages relative to the OAM measurement. The reconstructed Bell-state fidelities are $93.6(4.8)\%$, $92.9(5.9)\%$, $92.2(5.0)\%$, and $93.7(5.0)\%$ for $|\Phi^{+}\rangle$, $|\Phi^{-}\rangle$, $|\Psi^{+}\rangle$, and $|\Psi^{-}\rangle$, respectively. The corresponding Clauser--Horne--Shimony--Holt (CHSH) parameters are $S=2.34(12)$, $2.54(13)$, $2.46(12)$, and $2.42(11)$, all exceeding the classical bound $S\le2$ and confirming that nonlocal quantum correlations survive the OAM-to-polarization conversion.

A quantitative assessment of coherence transfer is obtained by computing the mutual fidelity between the reconstructed OAM state in Fig.~\ref{fig:OAM} and the corresponding polarization state for $|\Phi^{-}\rangle$ in Fig.~\ref{fig:Polarization}. The resulting value of $96.7\%$ with a statistical uncertainty at the few-percent level reflects the projection accuracy associated with the additional etalons and SLM settings used in the polarization measurement. Because the OAM tomography employs two etalons while the polarization tomography employs four, a simple estimate based on the product of the OAM fidelity and the mutual fidelity gives $0.964\times0.967\approx0.932$. Within the quoted error bars from the Monte Carlo analysis (see Supplemental Material), this estimate is consistent with the measured polarization fidelity of $92.9(5.9)\%$. This agreement suggests that the observed reduction of the Bell-state fidelities is compatible with the additional filtering and alignment imperfections introduced in the four-etalon configuration. At our present level of statistical precision, we do not resolve any extra decoherence that can be attributed to the OAM-to-polarization mapping itself.

The deviation of the measured OAM fidelity from unity is consistent with imperfect matching between the diffracted hypergeometric--Gaussian modes \cite{HyGGmode} and the Gaussian collection mode, together with ten-micron-level misalignment between the beam centers and the SLM phase singularities. The additional reduction in the polarization measurement is likewise compatible with the cumulative effect of two extra etalons and residual differential phase fluctuations between the spatial paths. Slow thermal drifts are compensated by reoptimizing the relative phase every fifteen minutes, and no systematic instability is observed on this timescale. Any faster fluctuations due to acoustic or vibrational noise are therefore below the sensitivity of the present measurements. Given the $2$--$6\%$ statistical uncertainties extracted from the Monte Carlo analysis, our data do not allow us to unambiguously separate these technical contributions from any smaller intrinsic decoherence mechanisms. They are, however, fully consistent with a scenario in which technical mode-projection and phase-stability limitations dominate the current infidelity budget.



In summary, we demonstrate a practical interface that converts OAM-entangled biphotons from a cold-atom double-$\Lambda$ system into polarization-entangled photons compatible with long-distance quantum networks. By coherently projecting a selected two-dimensional OAM subspace onto the polarization basis, we generate all four Bell states with fidelities of $92$--$94\%$, and an average CHSH value of $S=2.44$ confirms preserved nonlocality. The OAM-to-polarization mapping introduces no observable decoherence beyond technical limits associated with mode projection and phase stability. The same mechanism naturally extends to higher-dimensional photonic encodings. When combined with temporal gating or interferometric delays, the SLM-based mode-selective mapping can transcribe different OAM components into well-defined time bins, providing a pathway toward high-dimensional time-bin entanglement as reported in Ref.~\cite{timebin}. To the best of our knowledge, this work presents the first realization of OAM-to-polarization entanglement transfer in a cold-atom SFWM platform and establishes a viable building block for hybrid quantum networks that integrate atomic OAM resources with polarization-based architectures.



\textit{Acknowledgments}—This work was supported by the National Science and Technology Council of Taiwan under Grant Nos.~114-2112-M-006-007 and 114-2119-M-007-012. We also acknowledge support from the Center for Quantum Science and Technology within the framework of the Higher Education Sprout Project by the Ministry of Education in Taiwan.



\textit{Data availability}—The data that support the findings of this article are not publicly available. The data are available from the authors upon reasonable request.





\onecolumngrid
\newpage

\begin{center}
	\fontsize{12}{14.5}\selectfont
	\textbf{Supplemental Material for\\
		Polarization Entanglement in Atomic Biphotons via OAM-to-Spin Mapping}
	\\
	\fontsize{13}{10}\selectfont$\vphantom{1}$
	\\
	\fontsize{10}{12.5}\selectfont
	Chang-Wei Lin,$^1$ Yi-Ting Ma,$^1$ Jiun-Shiuan Shiu,$^{1,2}$ and Yong-Fan Chen$^{1,2}$\\
	\fontsize{9.2}{10.5}\selectfont
	\textit{
		$^1$Department of Physics, National Cheng Kung University, Tainan 70101, Taiwan\\ 
		$^2$Center for Quantum Frontiers of Research \& Technology, Tainan 70101, Taiwan}
	\\
	\fontsize{35}{10}\selectfont$\vphantom{1}$
	\\

\end{center}


\maketitle

\newcommand{\FigSOne}{
\begin{figure}[t]
	\centering
	\includegraphics[width = 17.6 cm]{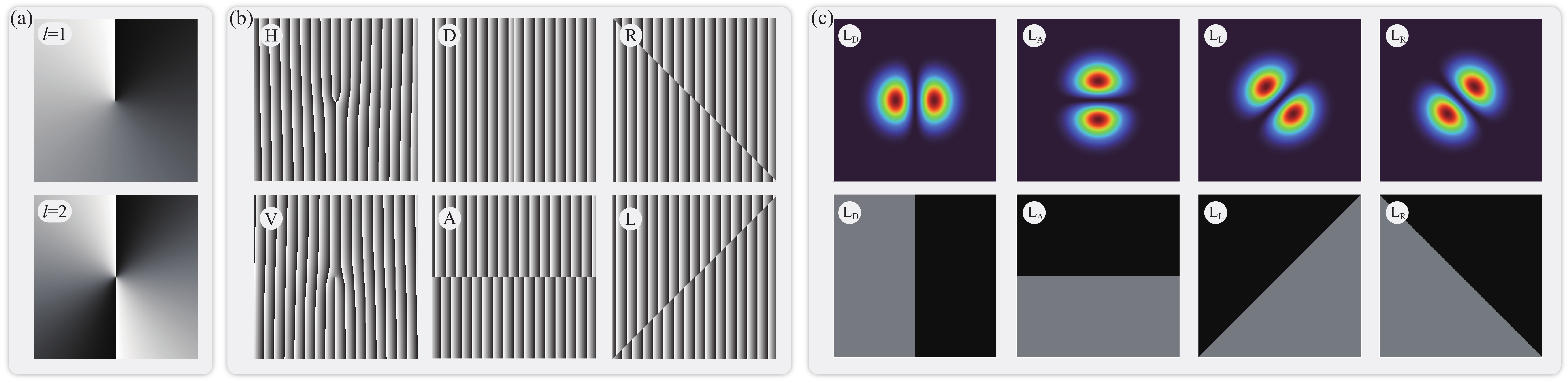}
	\caption{
		(a) Phase profiles of spiral phase plates (SPPs). The top panel shows the phase pattern for a topological charge of $l=1$, as used in the experiment, while the bottom panel displays an $l=2$ example for comparison. The grayscale colormap indicates the phase modulation from 0 (black) to $2\pi$ (white). (b) The six computer-generated holograms used for OAM quantum-state tomography; each index corresponds to the measurement bases defined in the text. (c) Theoretical spatial distributions of the optical fields for the four superposition projection bases ($|D\rangle$, $|A\rangle$, $|L\rangle$, $|R\rangle$). Intensity profiles are shown in the upper row and the corresponding phase distributions in the lower row. The sharp $\pi$ phase discontinuity across the nodal line is characteristic of these superposition modes.
		    }
	\label{fig:OAM1}
\end{figure}
}

\setcounter{equation}{0}
\setcounter{figure}{0}
\setcounter{table}{0}
\setcounter{page}{1}
\thispagestyle{empty} 
\makeatletter
\renewcommand{\theequation}{S\arabic{equation}}
\renewcommand{\thefigure}{S\arabic{figure}}
\renewcommand{\thetable}{S\arabic{table}}
\renewcommand{\bibnumfmt}[1]{[S#1]}


\section{Computer-generated holograms} \label{Pattern}

This section describes the computer-generated holograms implemented on the spatial light modulators (SLMs). These holograms support both the OAM-to-polarization conversion interface and the characterization of the OAM states. The discussion is divided into two parts. First, we outline the holographic grating used to map the entangled OAM modes onto the polarization degree of freedom, corresponding to Figs.~1(c) and 1(d) of the main text. Second, we present the six projective holograms used for quantum-state tomography in the OAM basis, as shown in Fig.~2(a) of the main text.

We begin by constructing the six holographic patterns used for OAM quantum-state tomography. The OAM degree of freedom is associated with an azimuthal phase structure that can be modeled as transmission through a spiral phase plate (SPP). As shown in Fig.~\ref{fig:OAM1}(a), an SPP with topological charge $l'$ imparts a phase modulation $\exp(i l' \phi)$ onto the incident wavefront and shifts its azimuthal quantum number by $l'$.

To implement this transformation on an SLM and to separate the modulated light from the unmodulated zero-order reflection, we superimpose a blazed grating phase profile onto the spiral phase. The grating function is
\begin{equation}
	\exp\!\bigl[i\,2\pi\,\mathrm{mod}(x/g,1)\bigr],
\end{equation}
where $g$ is the grating period and $x$ is the transverse coordinate. The blazed grating directs the modulated field into the first diffraction order, which is angularly separated from the zero order and can therefore be spatially isolated.

The hologram displayed on the SLM is the product of the spiral phase and the blazed grating,
\begin{equation}
	\exp\!\bigl[i\,l'\phi + i\,2\pi\,\mathrm{mod}(x/g,1)\bigr].
\end{equation}
The optical field diffracted into the first order acquires an OAM shift of $l'$. An incident $\mathrm{LG}^{\,l=-l'}_{0}$ mode is thereby converted into the fundamental Gaussian mode $\mathrm{LG}^{0}_{0}$, which can be efficiently coupled into a single-mode spatial filter and thus realizes a projective measurement.

The holograms for the two OAM eigenstate projections used in tomography,
$L_H \equiv |l=1\rangle$ and $L_V \equiv |l=-1\rangle$ [Fig.~\ref{fig:OAM1}(b)], are
\begin{align}
	\mathrm{Pattern}(L_H) &= \exp\!\bigl[-i\phi + i\,2\pi\,\mathrm{mod}(x/g,1)\bigr], \\
	\mathrm{Pattern}(L_V) &= \exp\!\bigl[i\phi + i\,2\pi\,\mathrm{mod}(x/g,1)\bigr].
\end{align}

The remaining four measurement bases are coherent superpositions of $|l=1\rangle$ and $|l=-1\rangle$,
\begin{align}
	L_D &\equiv |l=1\rangle + |l=-1\rangle, \\
	L_A &\equiv |l=1\rangle - |l=-1\rangle, \\
	L_L &\equiv |l=1\rangle + i|l=-1\rangle, \\
	L_R &\equiv |l=1\rangle - i|l=-1\rangle.
\end{align}
To project onto these superposition bases, the corresponding fields must again be transformed into the $\mathrm{LG}^{0}_{0}$ mode. As shown in Fig.~\ref{fig:OAM1}(c), the phase profiles of these superposition states are no longer helical; instead, they exhibit a binary phase structure with a $\pi$ phase step across the nodal line. These profiles are equivalent to first-order Hermite--Gaussian (HG) modes. The inverse transformation to $\mathrm{LG}^{0}_{0}$ is implemented by applying the conjugate binary phase pattern that flattens the wavefront. Superimposing this binary phase with the same blazed grating yields the holograms
\begin{align}
	\mathrm{Pattern}(L_D) &= \exp\!\bigl[i\pi\,\Theta(x) + i\,2\pi\,\mathrm{mod}(x/g,1)\bigr], \\
	\mathrm{Pattern}(L_A) &= \exp\!\bigl[i\pi\,\Theta(y) + i\,2\pi\,\mathrm{mod}(x/g,1)\bigr], \\
	\mathrm{Pattern}(L_L) &= \exp\!\bigl[i\pi\,\Theta(x+y) + i\,2\pi\,\mathrm{mod}(x/g,1)\bigr], \\
	\mathrm{Pattern}(L_R) &= \exp\!\bigl[i\pi\,\Theta(x-y) + i\,2\pi\,\mathrm{mod}(x/g,1)\bigr],
\end{align}
where $\Theta(\cdot)$ is the unit step function, equal to 1 for nonnegative arguments and 0 otherwise.  
These six holograms constitute a tomographically complete set for the OAM qubit subspace $\{|l=1\rangle, |l=-1\rangle\}$.

\FigSOne

We now turn to the holographic pattern used for OAM-to-polarization entanglement conversion. The goal is to design a single hologram that imparts $l=1$ to the $+1$ diffraction order and $l=-1$ to the $-1$ order simultaneously. The phase modulation required to add OAM $l$ in the $+1$ order is
\begin{equation}
	\exp\!\bigl[i\phi + i\,2\pi\,\mathrm{mod}(x/g,1)\bigr],
\end{equation}
whereas the modulation for the $-1$ order with OAM $-l$ is
\begin{equation}
	\exp\!\bigl[-i\phi - i\,2\pi\,\mathrm{mod}(x/g,1)\bigr].
\end{equation}
The target field at the SLM plane is therefore
\begin{align}
	E_{\mathrm{tar}}
	&= \exp\!\bigl[i\phi + i\,2\pi\,\mathrm{mod}(x/g,1)\bigr]
	+ \exp\!\bigl[-i\phi - i\,2\pi\,\mathrm{mod}(x/g,1)\bigr]  \nonumber \\
	&= 2\cos\!\bigl(\phi + 2\pi\,\mathrm{mod}(x/g,1)\bigr).
\end{align}
Because a phase-only SLM cannot directly encode continuous amplitude modulation, we normalize this field and retain only its sign, yielding the binarized pattern
\begin{equation}
	\mathrm{Pattern}(L_{\mathrm{tomo}})
	= \frac{E_{\mathrm{tar}}}{|E_{\mathrm{tar}}|}
	= \mathrm{sgn}\!\left[\cos\!\bigl(\phi + 2\pi\,\mathrm{mod}(x/g,1)\bigr)\right].
\end{equation}

The validity of this binarization follows from the Fourier expansion of a binary square wave,
\begin{equation}
	\mathrm{sgn}[\cos(\theta)]
	= \frac{4}{\pi}\!\left(\cos\theta - \frac{1}{3}\cos 3\theta + \frac{1}{5}\cos 5\theta - \dots\right).
\end{equation}
The fundamental harmonic reproduces the target cosine term, while the higher harmonics correspond to diffraction orders at larger angles ($3\theta, 5\theta,\dots$). Because our collection optics accept only the first diffraction order, these higher-order contributions are rejected. Consequently, the effective modulation experienced by the collected photons is equivalent to the ideal target pattern, apart from an overall diffraction efficiency. The complementary hologram used in the experiment is obtained by rotating this pattern by $180^\circ$ in the SLM plane.

\section{Theory of quantum tomography}

In this section we summarize the formalism used to reconstruct the bipartite density matrix from the projection measurements.

For the OAM degree of freedom, spatial filtering by the SLM and the single-mode filter reduces the biphoton state to an effective four-dimensional Hilbert space spanned by
\begin{equation}
	\bigl\{|1,1\rangle_L,\; |1,-1\rangle_L,\; |-1,1\rangle_L,\; |-1,-1\rangle_L\bigr\},
\end{equation}
where the subscript $L$ denotes the OAM Hilbert space and we use the shorthand $|l_s, l_{as}\rangle$. The polarization state is represented in an analogous basis
\begin{equation}
	\bigl\{|HH\rangle_P,\; |HV\rangle_P,\; |VH\rangle_P,\; |VV\rangle_P\bigr\},
\end{equation}
where $P$ denotes the polarization Hilbert space.

Because both sectors describe bipartite qubit states, they are mathematically isomorphic. For notational convenience we introduce the correspondence
\begin{equation}
	|l = 1\rangle \longleftrightarrow |H\rangle, \qquad
	|l = -1\rangle \longleftrightarrow |V\rangle,
\end{equation}
and express both OAM and polarization states in the logical basis $\{|H\rangle, |V\rangle\}$. The biphoton state is then fully described by a $4\times 4$ density matrix $\hat{\rho}$. In the computational basis $\{|HH\rangle, |HV\rangle, |VH\rangle, |VV\rangle\}$, it takes the form
\begin{equation}
	\hat{\rho} =
	\begin{pmatrix}
		\rho_{HH,HH} & \rho_{HH,HV} & \rho_{HH,VH} & \rho_{HH,VV} \\
		\rho_{HV,HH} & \rho_{HV,HV} & \rho_{HV,VH} & \rho_{HV,VV} \\
		\rho_{VH,HH} & \rho_{VH,HV} & \rho_{VH,VH} & \rho_{VH,VV} \\
		\rho_{VV,HH} & \rho_{VV,HV} & \rho_{VV,VH} & \rho_{VV,VV}
	\end{pmatrix}
	\equiv
	\begin{pmatrix}
		\rho_{11} & \rho_{12} & \rho_{13} & \rho_{14} \\
		\rho_{21} & \rho_{22} & \rho_{23} & \rho_{24} \\
		\rho_{31} & \rho_{32} & \rho_{33} & \rho_{34} \\
		\rho_{41} & \rho_{42} & \rho_{43} & \rho_{44}
	\end{pmatrix}.
\end{equation}

For a projection onto a two-photon state $|\psi\rangle$, the corresponding probability is
\begin{equation}
	P_\psi = \langle \psi | \hat{\rho} | \psi \rangle.
\end{equation}
A tomographically complete measurement set requires projections not only in the computational basis $\{|H\rangle, |V\rangle\}$ but also in their coherent superpositions,
\begin{align}
	|D\rangle &= \frac{|H\rangle + |V\rangle}{\sqrt{2}}, \qquad
	|A\rangle = \frac{|H\rangle - |V\rangle}{\sqrt{2}}, \\
	|L\rangle &= \frac{|H\rangle + i|V\rangle}{\sqrt{2}}, \qquad
	|R\rangle = \frac{|H\rangle - i|V\rangle}{\sqrt{2}} .
\end{align}

The diagonal elements (populations) are obtained directly from projections onto the computational basis,
\begin{align}
	\langle HH|\hat{\rho}|HH\rangle &= \rho_{11}, \\
	\langle HV|\hat{\rho}|HV\rangle &= \rho_{22}, \\
	\langle VH|\hat{\rho}|VH\rangle &= \rho_{33}, \\
	\langle VV|\hat{\rho}|VV\rangle &= \rho_{44}.
\end{align}
The off-diagonal elements are obtained from projections in the superposition bases. Using the sixteen joint projection settings summarized in Fig.~2(c) of the main text, we construct linear relations between the measured probabilities and the corresponding density-matrix elements. The real and imaginary parts of the off-diagonal entries follow as
\begin{align}
	\mathrm{Re}(\rho_{21}) &= \langle HD|\hat{\rho}|HD\rangle - \tfrac{1}{2}(\rho_{11} + \rho_{22}), \\
	\mathrm{Im}(\rho_{21}) &= \langle HL|\hat{\rho}|HL\rangle - \tfrac{1}{2}(\rho_{11} + \rho_{22}), \\
	\mathrm{Re}(\rho_{43}) &= \langle VD|\hat{\rho}|VD\rangle - \tfrac{1}{2}(\rho_{33} + \rho_{44}), \\
	\mathrm{Im}(\rho_{43}) &= \langle VL|\hat{\rho}|VL\rangle - \tfrac{1}{2}(\rho_{33} + \rho_{44}), \\
	\mathrm{Re}(\rho_{31}) &= \langle DH|\hat{\rho}|DH\rangle - \tfrac{1}{2}(\rho_{11} + \rho_{33}), \\
	\mathrm{Im}(\rho_{31}) &= \langle LH|\hat{\rho}|LH\rangle - \tfrac{1}{2}(\rho_{11} + \rho_{33}), \\
	\mathrm{Re}(\rho_{42}) &= \langle DV|\hat{\rho}|DV\rangle - \tfrac{1}{2}(\rho_{22} + \rho_{44}), \\
	\mathrm{Im}(\rho_{42}) &= \langle LV|\hat{\rho}|LV\rangle - \tfrac{1}{2}(\rho_{22} + \rho_{44}),
\end{align}
and
\begin{align}
	\mathrm{Re}(\rho_{41}) + \mathrm{Re}(\rho_{32}) 
	&= \langle DD|\hat{\rho}|DD\rangle 
	- \tfrac{1}{4}(\rho_{11} + \rho_{22} + \rho_{33} + \rho_{44}) \nonumber \\
	&\quad - \mathrm{Re}(\rho_{21}) - \mathrm{Re}(\rho_{31})
	- \mathrm{Im}(\rho_{42}) + \mathrm{Im}(\rho_{43}), \\
	\mathrm{Re}(\rho_{41}) - \mathrm{Re}(\rho_{32}) 
	&= \langle LR|\hat{\rho}|LR\rangle 
	- \tfrac{1}{4}(\rho_{11} + \rho_{22} + \rho_{33} + \rho_{44}) \nonumber \\
	&\quad + \mathrm{Im}(\rho_{21}) - \mathrm{Im}(\rho_{31})
	- \mathrm{Re}(\rho_{42}) - \mathrm{Re}(\rho_{43}), \\
	\mathrm{Im}(\rho_{41}) + \mathrm{Im}(\rho_{23}) 
	&= \langle RA|\hat{\rho}|RA\rangle 
	- \tfrac{1}{4}(\rho_{11} + \rho_{22} + \rho_{33} + \rho_{44}) \nonumber \\
	&\quad + \mathrm{Re}(\rho_{21}) + \mathrm{Im}(\rho_{31})
	+ \mathrm{Im}(\rho_{42}) + \mathrm{Re}(\rho_{43}), \\
	\mathrm{Im}(\rho_{41}) - \mathrm{Im}(\rho_{23}) 
	&= \langle AR|\hat{\rho}|AR\rangle 
	- \tfrac{1}{4}(\rho_{11} + \rho_{22} + \rho_{33} + \rho_{44}) \nonumber \\
	&\quad + \mathrm{Im}(\rho_{21}) + \mathrm{Re}(\rho_{31})
	+ \mathrm{Re}(\rho_{42}) + \mathrm{Im}(\rho_{43}).
\end{align}
These four coupled equations are evaluated using the previously determined populations and coherences. The full density matrix is then obtained by direct linear inversion, and a refined estimate is provided in the section on maximum-likelihood reconstruction, where physicality is enforced.

\section{Experimental implementation of tomography}

In the experiment we do not measure probabilities directly but instead record temporal cross-correlation histograms (coincidence counts). These must be converted into projection probabilities while accounting for channel-dependent detection efficiencies.

For each joint projection basis $|i,j\rangle$ with $i,j \in \{H,V,D,A,L,R\}$, we record the temporal waveform $G^{(\mathrm{exp})}_{ij}(\tau)$. The correlated coincidence contribution is obtained by integrating the peak of the histogram above the accidental background level. The background $G^{(\mathrm{exp})}_{ij}(\infty)$ is determined from the flat baseline at large time delays and corresponds to uncorrelated photon events.

To remove the effect of channel-dependent losses, we normalize the integrated signal by the intrinsic accidental rate of the source. For a fixed biphoton generation rate, the signal-to-background ratio is invariant under changes in the overall collection efficiency because both the correlated signal and the accidental background scale with the product of the detection efficiencies. Normalizing by the background level therefore cancels efficiency-dependent factors.

After subtracting the residual background due to stray light and detector dark counts, $G^{(\mathrm{env})}_{ij}(\infty)$, we define the normalized coincidence quantity
\begin{equation}
	\mathcal{G}_{ij}
	= \frac{\displaystyle\int_{0}^{T}
		\bigl[G^{(\mathrm{exp})}_{ij}(\tau)
		- G^{(\mathrm{exp})}_{ij}(\infty)\bigr]\, d\tau}
	{\displaystyle G^{(\mathrm{exp})}_{ij}(\infty)
		- G^{(\mathrm{env})}_{ij}(\infty)},
\end{equation}
where $T$ is the integration window covering the biphoton wave packet. The numerator gives the net correlated photon number, while the denominator removes the common efficiency factor.

To obtain physical probabilities we impose $\mathrm{Tr}(\hat{\rho})=1$.  
The computational-basis populations satisfy
$\rho_{HH,HH}+\rho_{HV,HV}+\rho_{VH,VH}+\rho_{VV,VV}=1$,  
so we identify
\begin{equation}
	\langle i,j|\hat{\rho}|i,j\rangle
	= \frac{\mathcal{G}_{ij}}
	{\mathcal{G}_{HH}+\mathcal{G}_{HV}+\mathcal{G}_{VH}+\mathcal{G}_{VV}} .
\end{equation}
These probabilities serve as the inputs to the linear inversion procedure of the preceding section.

\section{Maximum-likelihood reconstruction}

A physical density matrix $\hat{\rho}$ must satisfy three conditions, namely (i) normalization with $\mathrm{Tr}(\hat{\rho}) = 1$, (ii) Hermiticity with $\hat{\rho} = \hat{\rho}^\dagger$, and (iii) positive semidefiniteness with $\hat{\rho}\succeq 0$ such that all eigenvalues are nonnegative. The density matrix obtained from direct linear inversion satisfies the first two conditions by construction but may violate positive semidefiniteness due to statistical noise, resulting in small negative eigenvalues. To enforce physicality, we apply a maximum-likelihood estimation (MLE) procedure to obtain the physical state $\hat{\rho}_{\mathrm{MLE}}$ that ismost consistent with the measured data.

The MLE method assumes a known statistical model for the noise in the projection probabilities. For sufficiently large coincidence numbers, Poissonian statistics may be approximated as Gaussian, in which case maximizing the likelihood is equivalent to minimizing a weighted least-squares deviation between the measured probabilities and those predicted by $\hat{\rho}_{\mathrm{MLE}}$. Denoting the experimentally obtained probabilities as $P_\nu^{(\mathrm{exp})}$ and the theoretical probabilities for a trial state $\hat{\rho}$ as $P_\nu^{(\mathrm{th})}(\hat{\rho})=\langle \psi_\nu|\hat{\rho}|\psi_\nu\rangle$, we define the cost function
\begin{equation}
	\mathcal{L}
	= \sum_{\nu=1}^{16}
	\frac{ \bigl[P_\nu^{(\mathrm{exp})}
		- P_\nu^{(\mathrm{th})}(\hat{\rho}_{\mathrm{MLE}})\bigr]^2 }
	{2\sigma_\nu^2} ,
\end{equation}
where $\nu$ indexes the sixteen tomographic projectors, $|\psi_\nu\rangle$ are the corresponding two-photon states, and $\sigma_\nu$ are the standard deviations of the measured probabilities. The variances $\sigma_\nu$ are estimated from the coincidence counts using Poissonian statistics, $\sigma_\nu \propto \sqrt{N_\nu}$, where $N_\nu$ is the net coincidence count for projection $\nu$. The overall scale of $\sigma_\nu$ is irrelevant because only the minimum of $\mathcal{L}$ is used.

We numerically minimize $\mathcal{L}$ over the space of physical density matrices, enforcing normalization and positive semidefiniteness at each step. The resulting $\hat{\rho}_{\mathrm{MLE}}$ is then used to compute the fidelities and CHSH parameters reported in the main text.

\section{Error estimation}

Directly repeating the full set of sixteen tomographic measurements to obtain statistical uncertainties would require prohibitively long acquisition times. Instead, we estimate the uncertainties using a Monte Carlo procedure based on the Poissonian statistics of the coincidence counts.

For photon-counting measurements, the standard deviation of a measured count $N$ is $\sigma=\sqrt{N}$. Each experimentally obtained coincidence count $N_{\mathrm{exp}}$ is therefore modeled as a Gaussian random variable centered at $N_{\mathrm{exp}}$ with standard deviation $\sqrt{N_{\mathrm{exp}}}$. For each Bell-state configuration, we generate $10^4$ synthetic data sets, reconstruct the corresponding density matrix using the same procedure as for the experimental data, and compute the resulting state fidelities and CHSH parameters. The standard deviations of these distributions are taken as the one-standard-deviation statistical uncertainties of the reported values. Throughout this Supplemental Material, numbers in parentheses denote one-standard-deviation uncertainties in the last quoted digits; for example, $93.6(4.8)\%$ corresponds to $93.6\%\pm4.8\%$.

Applying this procedure, we obtain the following performance metrics. For $|\Phi^{+}\rangle$, the fidelity is $93.6(4.8)\%$ and the CHSH parameter is $S=2.34(12)$. For $|\Phi^{-}\rangle$, the fidelity is $92.9(5.9)\%$ with $S=2.54(13)$. For $|\Psi^{+}\rangle$, the fidelity is $92.2(5.0)\%$ with $S=2.46(12)$. For $|\Psi^{-}\rangle$, the fidelity is $93.7(5.0)\%$ with $S=2.42(11)$. The OAM-entangled state yields a fidelity of $96.4(2.2)\%$ and a CHSH violation of $S=2.77(6)$.

The uncertainties are larger than those commonly reported for SPDC-based sources. This difference arises primarily from the intrinsically low duty cycle of cold-atom SFWM systems, which limits the effective photon-generation window compared to continuous-wave SPDC. For comparable integration times and intrinsic pair-generation rates, the accumulated coincidence counts in our system are typically two orders of magnitude smaller than those of SPDC sources. Because statistical uncertainty scales as $1/\sqrt{N}$, the reduced counts impose a higher noise floor and thus lead to larger uncertainties in the reconstructed fidelities and CHSH parameters. The quoted uncertainties therefore largely reflect finite-count statistical noise rather than systematic drifts of the apparatus.


\end{document}